\documentclass[11pt]{article}
\date{} 

\usepackage[a4paper, top=1.8in, bottom=1.5in, left=1.4in, right=1.4in]{geometry}

\usepackage[T1]{fontenc}
\usepackage[utf8]{inputenc}
\usepackage{lmodern}
\usepackage{microtype}

\usepackage{amsmath,amssymb,amsthm}

\usepackage[dvipsnames,svgnames,x11names]{xcolor}
\usepackage{graphicx}

\usepackage{array}
\usepackage{booktabs}
\usepackage{longtable}

\usepackage{float}
\usepackage{caption}
\usepackage{subcaption}

\usepackage{xurl}   
\usepackage[pdfencoding=auto, psdextra]{hyperref}
\hypersetup{
	pdftitle={Predictive posteriors under hidden confounding},
	pdfauthor={Carlos García Meixide; David Ríos Insua},
	pdfkeywords={3 to 6 keywords, that do not appear in the title},
	colorlinks=true,
	linkcolor=blue,
	citecolor=Blue,
	urlcolor=Blue
}

\usepackage[authoryear,round]{natbib}
\newtheorem{ass}{Assumption}
\newtheorem{thm}{Theorem}

\newtheorem*{remark}{Remark}

\newcommand{\E}{\mathbb{E}}
\newcommand{\K}{\mathcal{K}}

\newcommand{\anon}{1}
\begin{document}

\def\spacingset#1{\renewcommand{\baselinestretch}%
{#1}\small\normalsize} \spacingset{1}


\if1\anon
{
  \title{\bf Predictive posteriors under hidden confounding}
  \author{Carlos García Meixide\\
    and \\
    David Ríos Insua \\}
  \maketitle
} \fi

\if0\anon
{
  \bigskip
  \bigskip
  \bigskip
  \begin{center}
    {\LARGE\bf Predictive posteriors under hidden confounding}
\end{center}
  \medskip
} \fi

\bigskip
\begin{abstract}
	Predicting outcomes in external domains is challenging due to hidden confounders that potentially influence both predictors and outcomes. Well-established methods frequently rely on stringent assumptions, explicit knowledge about the distribution shift across domains, or bias-inducing regularization schemes to enhance generalization. While recent developments in point prediction under hidden confounding attempt to mitigate these shortcomings, they generally do not provide principled uncertainty quantification. We introduce a Bayesian framework that yields well-calibrated predictive distributions across external domains, supports valid model inference, and achieves posterior contraction rates that improve as the number of observed datasets increases. Simulations and a medical application highlight the remarkable empirical coverage of our approach, nearly unchanged when transitioning from low- to moderate-dimensional settings.
\end{abstract}

\noindent%
\vfill

\newpage
\spacingset{1.8} 

	\section{Introduction}
	
	
	In recent years, there has been a surge of interest in machine learning (ML) methods that demonstrate robust performance even when there are shifts in the distribution between training and test data \citep{shiftllms}. When an ML algorithm is trained on a labeled source dataset, it can be expected to perform well on an unlabeled target dataset drawn from the \emph{same} distribution, provided that generalization guarantees hold. However, numerous applications often experience distribution shifts, challenging the standard iid assumption and the use of existing prediction methods. This paper focuses on delivering probabilistic predictions for a response variable \( Y \in \mathbb{R} \) based on a set of covariates \( X \in \mathbb{R}^p \), 
    exploiting the integration of heterogeneous data sources \cite{Zhaijcgs} so that well-calibrated predictive uncertainty quantification is achieved. 
	
	The majority of research addressing distribution shifts has concentrated on point prediction in regression or classification. Notable strategies include {\em distributionally robust optimization} \citep{sinha2020}, which involves minimax optimization over distributions; and the causality-inspired framework initiated by \cite{anchor}. However, their regularization scheme precludes asymptotically unbiased estimation of causal parameters. A promising framework for quantifying uncertainty of predictive algorithms is {\em conformal prediction}, which pursues the construction of robust prediction intervals for a fixed marginal coverage level based on order statistics \citep{tibs19}. In addition, causality has been related to a certain risk minimization problem entailing a new interpretation of robustness \citep{stspeter}. However, to our knowledge, only two studies address robust probabilistic prediction in the causality literature: \cite{kook} proposed {\em distributional anchor regression} as an extension for conditional transformation models; and \cite{henzi} analysed invariance for probabilistic predictions penalizing empirical risk minimization with the variance of the scoring rule across environments.

	Covariate shifts refer to differences in the marginal distribution of \( X \) between training and target environments, due e.g. to age or gender distributional structure variations. In turn, \( Y | X \) shifts involve changes in the conditional relationship between $Y$ and $X$, potentially due to unobserved confounders or noncompliance \citep{Robins1989}. \cite{ai2024} argue that distribution shifts comprise both ``$X$'' and ``\( Y | X \)'' types of shifts, which are fundamentally different: while the $X$ type can be identifiable from data, the type $Y | X$ itself is not inherently identifiable. 
    
    Work introduced in  \cite{meixide2025unsuperviseddomainadaptationhidden} enables identification under $Y | X$ shifts through Generative Invariance (GI), a projection estimand that adapts predictions to unseen domains without requiring knowledge of the specific test distribution shifts or hyperparameter tuning. It achieves identifiability under mild, verifiable conditions accomodating the influence of unobserved confounders across heterogeneous training environments through a parametric component, therefore delivering optimal predictions while preserving unbiased estimation of causal parameters. This marks a fundamental change from the established view of predictive robustness as a minimax optimization problem \citep{anchor}, as GI does not require introducing bias into estimators of causal parameters. It makes this possible through a novel mechanism that overpasses regularization techniques typically used in robustness frameworks \citep{anchor,shen2025}. Despite being the first method exploiting distributional information from unlabeled data in the target domain to provide asymptotically unbiased domain adaptation, GI cannot say anything about uncertainty at the predictive scale on $Y$, though. 
	
 To overcome this, the present paper provides principled credible intervals for causal parameters under hidden confounding and, therefore, enables simultaneous causal discovery and calibrated predictions on distributionally shifted test domains: to the best of our knowledge, we deliver the first computationally feasible hypothesis test of a covariate being in the parental set of a target response in a linear structural equation model. Note that obtaining credible statements for existing causal discovery methods is often difficult as one would need to determine the distribution of causal effects estimators after having searched and estimated a model graphical structure, as \cite{peters} and \cite{fan2024environment} attempts clearly showcase. Importantly, our method will calibrate well even when prior information is not available, as simulations in Section \ref{sec:sim} show.  The key message is that the prior is not used to incorporate subjective information—it simply absorbs the effects of weak identifiability \citep{neweak}. This is analogous to the well known example of Gaussian priors emerging as $l_2$ penalties in ridge regression: while not representing expert domain knowledge, they entail important technical consequences in regularization terms.
		\begin{figure}[t!]
		\centering
		\begin{minipage}{0.48\textwidth}
			\centering
			\includegraphics[width=\linewidth]{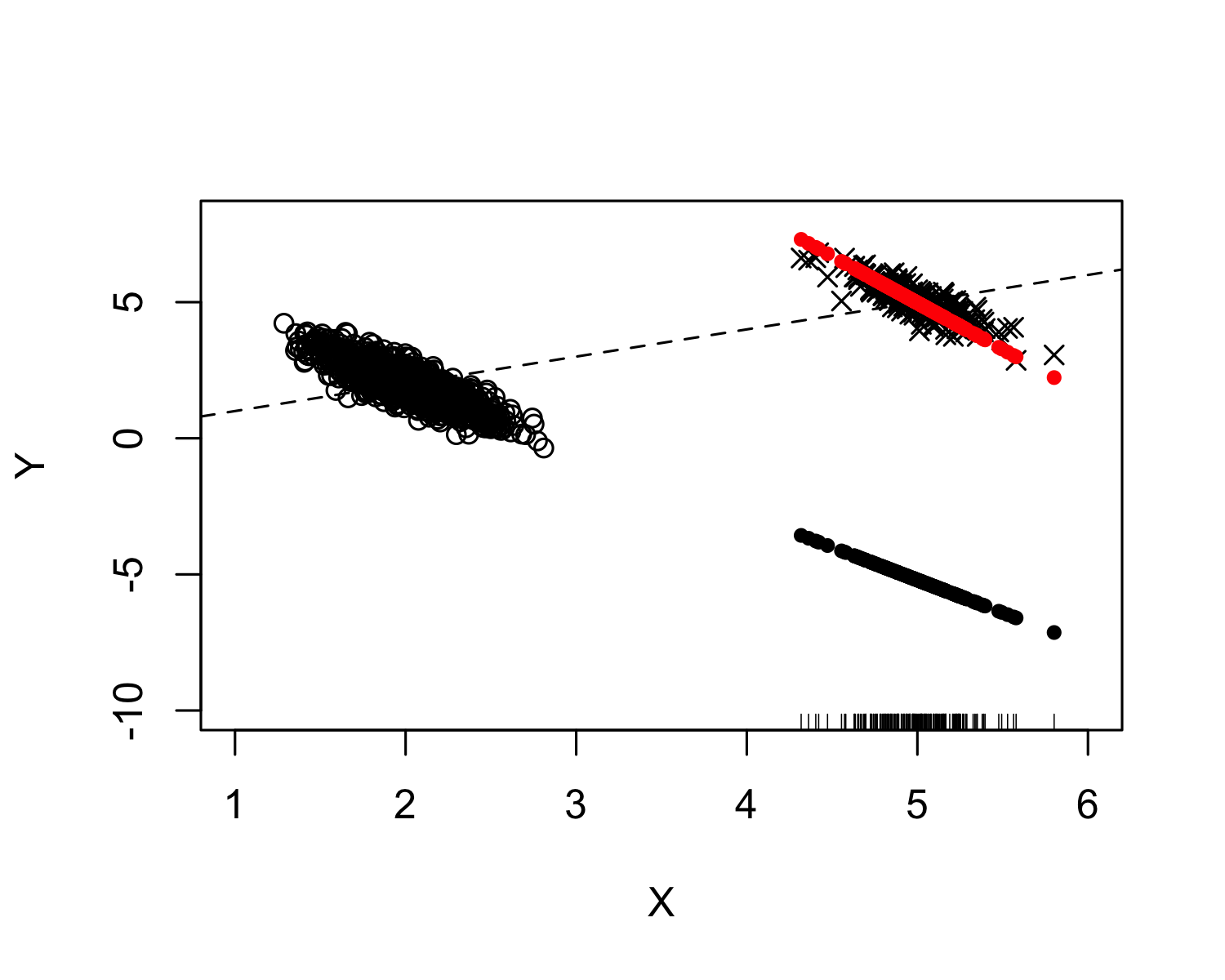}
			\caption{An OLS model fitted to the training data (circles) produces solid black prediction dots, falling completely off-target. In contrast, the red dots represent GI predictions.}
			\label{firstfig}
		\end{minipage}
		\hfill
		\begin{minipage}{0.48\textwidth}
			\centering
			\includegraphics[width=\linewidth]{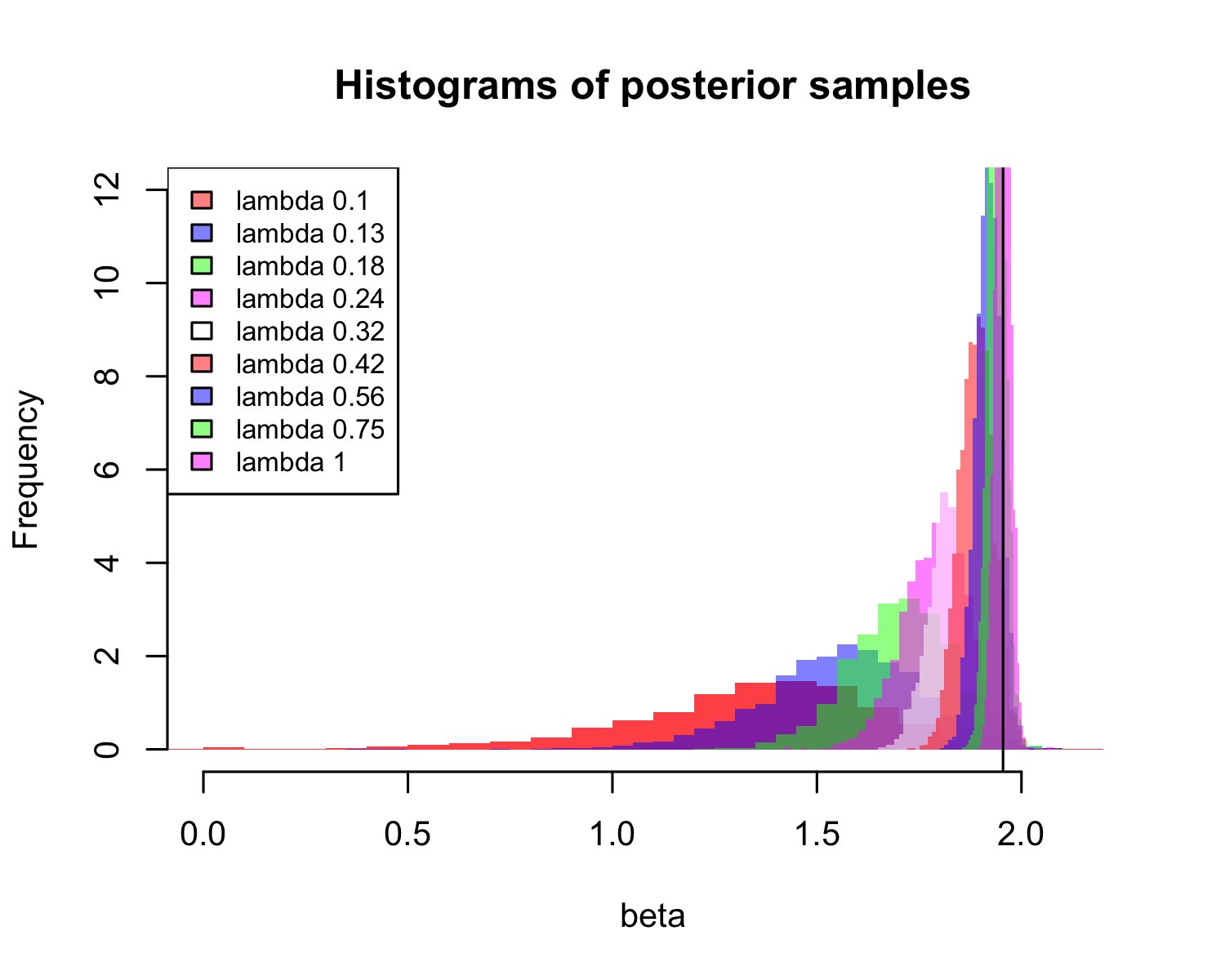}
			\caption{Prior influence on posterior of $\beta$ under weak identifiability indexed by $\lambda$.}
			\label{violation}
		\end{minipage}
	\end{figure}

         With this, we offer the first principled, unified, and practical solution to two longstanding and challenging problems in computational statistics: conducting hypothesis testing for causal parenting and calibrating probabilistic predictions in the presence of hidden confounding.
      In related work, \cite{wu} introduce a hierarchical generative model that encodes the indices of invariant predictors as a latent Bernoulli vector and, then perform posterior inference to select features whose conditional relationship with the outcome remains stable across environments. Nonetheless, their framework exploits the original invariance assumption in \cite{peters} and, therefore, it inherits its vulnerability: when hidden confounders induce violations of this assumption, the method can still select non-causal features or discard causal ones. Likewise,  \cite{madaleno2025} reformulate \cite{peters} in a Bayesian setting, yet it still relies on the invariance assumption. Finally, \cite{holovchak2025} seek to recover interventional distributions, which are not designed for predictive purposes and do not provide built-in facilities for formal hypothesis testing. Our approach, alternatively, supplies well-calibrated Bayesian posteriors for both causal parameters and outcome predictions, yielding credible intervals on a shifted test environment and a causal discovery procedure that remain valid under hidden confounding. Besides, we computationally assess the empirical coverage properties of our approach across a range of scenarios, varying both sample size and covariate dimension. In all settings, the coverage achieved by our method closely matches the nominal theoretical level. In contrast, even the best-case performance of the baseline—Ordinary Least Squares (OLS)—is comparable to the worst-case performance of our method.

	


	\section{Method}

	Let us recall the basic setup in \cite{meixide2025unsuperviseddomainadaptationhidden}
	to motivate the problem of interest.
	Consider analyzing the relationship between the random variables $X \in \mathbb{R}^p$ and $Y \in \mathbb{R}$, using the structural model
	\begin{equation} \label{scm_main}
		X \leftarrow \sum_z 1\{Z=z\}X_z \quad \textrm{and} \quad Y \leftarrow \alpha_* + \gamma^T_*X + \varepsilon_Y,
	\end{equation}
	where $Z \geq 0$ is a discrete indicator distinguishing between training ($Z \geq 1$) and testing ($Z=0$) regimes, with $X_z$ mutually independent $p$-dimensional random covariate vectors. $\alpha_* \in \mathbb{R}$ represents an intercept and $\gamma_* \in \mathbb{R}^p$ the vector of slopes. The error term $\varepsilon_Y \in \mathbb{R}$ satisfies $\mathbb{E}[\varepsilon_Y|Z] = 0$, but may exhibit dependence with $X_z$ due to hidden confounders. Our primary objective is to provide predictions for a test covariate sample ${ D}_0=\{ X_{0,1},\ldots,X_{0,{n_0}} \}$, corresponding to $Z=0$, after having observed samples involving \textit{both} response and covariates for $Z \geq 1$ . Note that model (\ref{scm_main}) does not impose any specific structure on the nature of the distribution shifts among the $X_z$'s (e.g., additive), unlike commonly assumed in the literature.

 For introductory purposes, assume a one dimensional covariate setting with $\alpha_*=0$. Theorem 3.2 in \cite{meixide2025unsuperviseddomainadaptationhidden} enables frequentist identification of the causal parameter $\gamma_*$ in (1) if $\mathbb{E}[X \mid Z = 1 ] = \mathbb{E}[X_1 ] := \mu \neq 0$ without requiring heterogeneous training sources,
     showing that the minimizer $(\gamma_{opt}, K_{opt})$ of the population least squares risk 
    $ (\gamma,K) \longmapsto \mathbb{E}\left[(Y-(\gamma X+K(X-\mathbb{E} X)))^2 \mid Z = 1\right]$
    exists, is unique and attained at $\left(\gamma_*, \frac{K_*}{\operatorname{Var}X_0}\right)$ with $K_*:=\operatorname{Cov}(\varepsilon_Y,X_0) = \operatorname{Cov}(\varepsilon_Y,X_1) $ under \textit{inner-product invariance} \citep{causaldan}. Therefore, the conditional expectation $Y_0 | X_0$ becomes identified despite the presence of hidden confounders. A $p$-dimensional version with intercept of this identifiability result is provided in Theorem 3.4 therein, stating that if the population mean vectors of the covariate distributions corresponding to different data environments span $\mathbb{R}^p$, then $\alpha_*, \gamma_*$ and $K_*$ are identifiable. 
   		
	Figure \ref{firstfig} illustrates potential issues of
	standard methods when handling this  problem. The empty circles ($n=1000$) serve for training purposes to fit a standard OLS linear model. Samples from a different test 
	environment ($n_0 = 200$) are reflected with crosses. 
	Predictions based on the trained model -the solid dots-  clearly underperform, as they miss the crosses. In turn, frequentist GI \citep{meixide2025unsuperviseddomainadaptationhidden} (red dots) is able to capture the structure of the test environments by just looking at the empty points and the rug on the horizontal axis, 
     moving away from traditional reliance on regularization and \textit{finite robustness} \citep{bental} to implement optimal predictions in unlabeled domains without sacrificing consistent estimation of causal parameters.

	We observe $E$ training environments, each with $n_e$ i.i.d.\ samples. Let
	\[
	(X_{e,1},Y_{e,1}),\ldots,(X_{e,n_e},Y_{e,n_e}) \sim \mathbb P_e,\qquad e=1,\ldots,E,
	\]
	and denote $N:=\sum_{e=1}^E n_e$. For notational convenience we stack all observations as $D_N=\{(Z_i,X_i,Y_i)\}_{i=1}^N$, where $Z_i\in\{1,\ldots,E\}$ indicates the environment of the $i$th sample. The full test dataset is $(X_{0,1},Y_{0,1}),\ldots,(X_{0,n_0},Y_{0,n_0}) \sim \mathbb P_0$ but only $X_{0,j}$’s are observed at prediction time. For each environment $e$, we build the likelihood upon the following hierarchical model
	\begin{equation}\label{eq:M}
		\begin{aligned}
			X_{ei} &\sim \mathcal N_p(\mu_e,\Sigma_e),\\
			Y_{ei}\mid X_{ei},w,\vartheta_e &\sim 
			\mathcal N\!\Big(\alpha +  \gamma^TX_{ei} + K^\top (X_{ei}-\mu_e),\ \sigma_{Y}^2\Big),
		\end{aligned}       
	\end{equation}
		The parameter of interest is
	$
	w := (\beta, K)\in \mathbb R^{2p+1},
	$
	where
 \( \beta= (\alpha,\gamma) \in \mathbb R^{p+1} \) contains the regression coefficients $\gamma$ and the intercept $\alpha$, and
 \( K \in \mathbb R^p \) is the vector associated with the correction term absorbing the effect of hidden confounders. In addition,  $\vartheta_e:=(\mu_e,\Sigma_e,\sigma_{Y}^2), e = 1,\ldots, E$ are the environment–specific nuisances. Technically, the original $p-$dimensional covariates are augmented with a unit entry at the beginning to include an intercept, so \( \beta \) has dimension \( p+1 \). The 
  term \( K^\top(X_{ei} - \mu_e) \) only acts on the original \( p \) covariates (not the intercept), so \( K \in \mathbb R^p \). Hence, the total parameter dimension is \( (p+1) + p = 2p + 1 \), and the full parameter vector \( w \) lies in \( \mathbb R^{2p+1} \). Note that such correction term remains completely unrelated to the intercept since
	$
	\mathbb E\!\left[K^\top(X-\mu_e)\mid Z=e\right]=
	K^\top\bigl(\mathbb E[X\mid Z=e]-\mu_e\bigr)=0.
	$
	
Let us adopt ridge–type Gaussian priors on $(\beta,K)$ \citep{tew} and weakly informative priors on the nuisance parameters:
\[
\begin{aligned}
	\mu_1,\ldots,\mu_E &\sim \mathcal N_p(\hat\mu,\Sigma_\mu),\\
    \alpha &\sim N(0,\tau^2\sigma^2_Y) \\
	(\gamma,K)\mid \tau^2,\sigma_Y^2 &\sim \mathcal N_{2p}\bigl(0,\ \tau^2\sigma_Y^2 I_{2p}\bigr),\\
	\sigma_Y &\sim \pi(\sigma_Y) \propto \frac{1}{\sigma_Y},\quad (\sigma_Y > 0) \quad & \\
	\tau^2 &\sim \text{Beta-prime}(a_\tau,b_\tau),
\end{aligned}
\]
where $\hat \mu = \frac{1}{N}\sum_{i=1}^N X_i$. The prior on $\sigma_Y$ is the standard noninformative Jeffreys prior for scale parameters.	In addition, $\Sigma_\mu$ is built upon a correlation matrix $R$ with prior distribution $R \sim \text{LKJ}(2)$ \citep{lewandowski2009generating}, with  single shape parameter $\eta$ so that the probability density function for a $p \times p$ matrix $R$ is $
\propto [\operatorname{det}({R})]^{\eta-1}
$.  The Beta–prime distribution \citep{polson} has density $\pi(\tau^2)=\frac{(\tau^2)^{a_\tau-1}(1+\tau^2)^{-a_\tau-b_\tau}}{B(a_\tau,b_\tau)},
	$
	where $B$ represents the Beta function, with the common choice $a_\tau=b_\tau=1/2$ yielding a standard half–Cauchy on $\tau$.   The latter is particularly useful as a weakly informative prior for scale parameters because of its heavy tails. 
Observe though that specification of priors is not the focus of this work. Section~\ref{sec:theo} actually shows that the posterior contraction rate depends on the prior only through multiplicative constants; the exponential rate is driven entirely by the likelihood, not by prior tails.

	In the canonical Bayesian linear model, the posterior mode (and mean) is equivalent to the frequentist ridge estimate with penalty $1 / \tau^2$. In our case, we  express the population heterogeneity leading to each environment as sampled from a common prior distribution to get a beneficial shrinkage effect.  $N_p(\hat{\mu}, {\Sigma_\mu})$ is 
     thus viewed in our setup as an empirical Bayes distribution characterizing the variability of the covariate means induced by naturally occurring environments.  It is useful to think of the first step of the data generating process (\ref{eq:M}) as the covariate means being drawn from a common prior. Note though that we just plug the empirical mean and not the variance of this distribution, to avoid the following recurrent problem in empirical Bayes estimation \citep{ritov}. For simplicity, assume that $p=1$ and all the environments gave rise to the same number  $n$ of observations. Then, it is possible to build, via maximum likelihood, an estimator for the Gaussian prior variance as $  \max \left\{\frac{1}{E} \sum_{e=1}^E\left(\hat{\mu}_e-\hat{\mu}\right)^2-\frac{\hat{\sigma}^2}{n}, 0\right\}$
	where $\hat{\sigma}$ is a consistent estimator of an asymptotic variance and $\hat \mu_e = \frac{1}{n_e}\sum_{i=1}^{n_e} X_{ei}$. If the previous max collapses
     to 0, then the resulting posterior is a point mass and the empirical Bayes posterior predictive region is a singleton. We escape from this problem by associating a hyperprior to this parameter. 
     
     To avoid specifying additional priors for the covariance matrices $\Sigma_1, \ldots, \Sigma_E$, we estimate each $\Sigma_e$ empirically from the data in the $e-$th environment and plug these estimates in the likelihood (\ref{eq:M}). The estimation error is asymptotically negligible and does not affect posterior behavior of the parameters  as we next see. For the sample covariance $\widehat\Sigma_e$ of $n_e$ i.i.d.\ Gaussian observations with population covariance $\Sigma_e$, the following high–probability bound holds \citep{Chandrasekaran2012}: let $\psi_e=\|\Sigma_e\|_2$, fix any $\delta\in(0,8\psi_e]$, and suppose
     $n_e \ge 64\,p\,\psi_e^2/\delta^2$. Then $\mathbb{P}\!\bigl(\|\widehat\Sigma_e-\Sigma_e\|_2 \ge \delta\bigr)
     \;\le\; 2\exp\!\Bigl(-\,\frac{n_e\delta^2}{128\,\psi_e^2}\Bigr)$. Thus, choosing $\delta\simeq C\,\psi_e\sqrt{p/n_e}$ gives the familiar $\|\widehat\Sigma_e-\Sigma_e\|_2=O_{\mathbb P}\!\big(\psi_e\sqrt{p/n_e}\big)$ rate (for fixed $p$, this reduces to $n_e^{-1/2}$, but with growing $p$ the $\sqrt{p/n_e}$ factor would be explicit). This is the rate we reference in Assumption \ref{ass:LAN} later on. 
     
      It is important to emphasize the distinction between this procedure for the environments' covariance structure and the handling of the environment means $\mu_1, \ldots, \mu_E$ of the covariates: these are not estimated and plugged in. Instead, they are treated as unknown parameters with a proper hierarchical prior of the form $\mu_e \sim \mathcal N_p(\hat\mu, \Sigma_\mu)$, where only the prior mean $\hat\mu$ is empirically estimated from the pooled covariate data. That is, the centering point of the prior is data-dependent, but the $\mu_e$’s themselves are fully inferred under the model. This distinction ensures a coherent Bayesian treatment of the $\mu_e$’s while maintaining computational simplicity for the $\Sigma_e$’s. This completes our formulation above as deliberately compatible with the regularity conditions required for Bernstein–von Mises results stated in the asymptotics, Section \ref{sec:theo}. 

Let us now describe how to obtain the predictive distribution of $Y_0 $ conditional on $X_0$. \cite{meixide2025unsuperviseddomainadaptationhidden} show that we recover the test distribution under inner product invariance \citep{causaldan}, error Gaussianity and the so-called \textit{causal Mahalanobis conditions} as
	$	(X_0 ,
	\alpha_*+	\gamma^T_*X_0 + K_*^T \Sigma_0^{-1}\left(X_0-\mathbb{E} X_0\right)+\xi S_0
) \sim \mathbb{P}_0
	$
	with $S_0= \sqrt{\sigma_Y^2-K_*^T \Sigma_0^{-1} K_*}$ and $\xi \sim N(0,1)$ independent of $X_0$. Then, the predictive distribution for $f = \mathbb{E}\left[Y_0 | X_0, w \right] = 	\alpha + \gamma^TX_0 + K^T \Sigma_0^{-1}\left(X_0-\mathbb{E} X_0\right)$ is given by averaging it with respect to the posterior $
	\begin{aligned}
		p\left(f \mid X_0,D_N\right) & =\int p\left(f \mid X_0, {w}\right) p({w} \mid D_N) d{w},
	\end{aligned}
	$ from which we obtain the predictive posterior by adding an independent noise term with variance $S_0$ coming from aleatoric uncertainty. 
	
	An example of a case where the predictive posterior has a closed form expression is given by trivially pooling all the observations into one single training source (denoted by ${D}$) without considering the dataset indicators and assuming a fixed-design linear model $f = X^T_0 w$ with additive Gaussian noise $Y = f + \varepsilon_Y$,  $\varepsilon_Y \sim N(0,\sigma^2_Y)$. Then the posterior distribution of the conditional mean is $p\left(f \mid X_0, D\right) = N\left(\frac{1}{\sigma_Y^2} X_0^{\top} A^{-1} X^T Y, X_0^{\top} A^{-1} X_0\right) $ with $A=\sigma_Y^{-2} X^T X+\Sigma_p^{-1}$ and $\Sigma_p$ being the prior covariance matrix of $w$ \citep{rasmussen}. The posterior predictive distribution would be $p\left(Y_0 \mid X_0,D \right) = N\left(\frac{1}{\sigma_n^2} X_0^{\top} A^{-1} X^T Y, X_0^{\top} A^{-1} X_0 + \sigma^2_Y\right)$ and we use it to build a credible interval $\hat{C}\left(\cdot\right)$ by considering, for instance, its $\frac{\alpha}{2}$ and $1 - \frac{\alpha}{2}$ quantiles. 
	
		We could derive the explicit expression for the conditional posterior distributions of $\beta$ and $K$ given ${D}_N$, as 
	we do in the Appendix, but it would not provide additional insights. Therefore the model is directly implemented in 
	\texttt{Rstan} \citep{guo2020package}. Even in our more complex case, where $f$ and $\varepsilon_Y$ are not independent, even conditional on the parameters, sampling from the conditional predictive distribution $p\left(Y_0 \mid X_0,D_N \right) $ is still a matter of creating a \texttt{generated quantities} block in \texttt{Rstan} preceded by the likelihood and prior specifications discussed in this section and then sampling \texttt{normal\_rng(cond\_mean, sigmay);} where \texttt{cond\_mean} is $f$ and \texttt{sigmay} is $\sigma_Y$. The computational properties of our estimator will be explored in Sections \ref{sec:sim} and \ref{sec:quiron}.


		\subsection{Causal discovery under hidden confounding}\label{sec:sel}
	
	A type I error guarantee \citep{peters} for causal discovery is known to hold under an \textit{invariance} hypothesis: the conditional distribution of the response given all direct causal predictors must remain identical under interventions on variables other than the target variable.  The authors consider the intersection $\mathcal{\hat S}$ of all the subsets of covariates not rejecting the null hypothesis of invariance to build a test capable of controlling the proportion of false positive selections: $\mathbb{P}[\hat{\mathcal{S}} \subseteq \mathrm{pa}(Y)] \geq 1-\alpha $ where $ \mathrm{pa}(Y) = \{1\leq j \leq p: \gamma^j\neq 0\}$ . Using the intersection makes their procedure conservative in terms of power, which, besides, depends highly on the  the observed environments structure. However, the presence of hidden variables (the ``$Y|X$'' shift) breaks invariance.
	
	 Our procedure correctly estimates the posterior distribution of the causal parameters $\beta$ and $K$, and, therefore, automatically enables for hypothesis testing of one component $\gamma^j$ being zero, $j = 1,...,p$: our method provides the posterior distribution $p(\gamma^j | D_N)$ for each coefficient, and we can construct a $(1-\alpha)$ credible region $\hat C_\alpha$ such that:
	$\mathbb{P}(\gamma^j \in \hat  C_\alpha | D_N) = 1-\alpha$. 
	Given $\hat C_\alpha$, we assess whether $\gamma^j = 0$ is plausible by checking whether $0$ lies within $\hat C_\alpha$.  Our solution to this problem is also more natural because the invariance hypothesis is always formulated based on the \textit{observed} environments $\mathcal{E}$: a subset of covariates $S \subset \{1,\ldots,p\}$ fulfills the null hypothesis $H_{0,S}(\mathcal{E})$- ``invariance''- iff  $\mathcal{L}\left(Y^e \mid X_S^e\right)$ (the conditional distribution of the response given the covariates in $S$) is the same for all $e \in \mathcal{E}$. This may seem a bit contradictory 
	as hypotheses should be assertions that hold or not pre-empirically, leading to a paradox: knowing which environments are observed or not turns out to be something known population-wise, prior to sampling-- what $\mathcal{E}$ is becomes an aspect of the underlying distribution before we even have sampled from it. We solve this issue automatically: given a sample $\gamma^j_1, \ldots ,\gamma^j_m$ from the posterior of $\gamma^j$ we decide that $j$ is a causal parent of $Y$ when $\min \left\{\left|\left\{i: \gamma^j_{{i}}<0\right\}\right|,\left|\left\{i: \gamma^j_{{i}}>0\right\}\right|\right\}<\alpha m,$
	interpreted as the posterior distribution being sufficiently shifted away from zero, either above or below.
	
	We acknowledge that classical Bayesian hypothesis testing often proceeds via explicit model selection priors (e.g., spike–and–slab \citep{ishwaran2005spike}) and Bayes factors \citep{safe}. Our primary goal here, however, is Bayesian decision support for causal parenting under hidden confounding, using the full posterior over $(\beta, K)$ induced by our method. Our rule—declaring $j$ a parent when the posterior mass on one half-line exceeds $1-\alpha$ —is a decision criterion that directly controls the posterior probability of a sign error. Thus, although the presentation resembles a frequentist test, the quantity thresholded is a posterior probability, not a $p$-value, and the operational criterion is minimization of posterior expected loss, not type-I error \citep{french2000statistical}.  A separate consideration is that our decisions inherit good frequentist properties because our regime admits a valid Bernstein–von Mises approximation as argued in Section 2.2.
	We also agree that model-selection priors provide a complementary “fully Bayesian” route. Our framework could incorporate them (e.g., spike–and–slab on $\gamma$, continuous shrinkage such as the horseshoe, or hierarchical slabs that couple $\gamma$ and $K$). We chose not to foreground these here for two reasons. First, under hidden confounding, any selection prior must be specified jointly on $(\gamma, K)$ to respect the dependence structure revealed by the explicit posterior in Section 2.2, since naïvely placing independent spikes on each $\gamma^j$ can lead to misleading inclusion probabilities. Second, scalable inference with high-dimensional discrete model spaces can be computationally demanding and sensitive to slab hyperparameters, whereas our posterior probability decision rule remains computationally simple and directly tied to predictive calibration. That said, we view selection priors as a valuable extension: one can report posterior inclusion probabilities and/or control the local false sign rate alongside our current decision rule. 
	

	\subsection{Asymptotics}\label{sec:theo}
	
	We develop the large sample theory for the posterior distribution arising from likelihood and prior especifications in Section 2, establishing posterior consistency, contraction rates and a Bernstein-von Mises type of theorem. Notably, our results highlight two key phenomena often left unexplained in frequentist theory: robustness to near-violations of identifiability, and the role of natural variability across environments in shaping asymptotic behavior. Proofs for all the results in this section can be found in Supplement A. Let \(D_{n,E}
		=\{(X_{ei},Y_{ei})\,:\,e=1,\dots,E,\;i=1,\dots,n\}\) be the different samples, 
		assumed i.i.d.\ within each
		environment. To streamline notation, we assume wlog the same sample size $n$ across environments. Let \(\ell_{ei}(w)=\log\!\bigl\{p(Y_{ei},X_{ei}\mid w^\star,\vartheta_e^\star)/
		p(Y_{ei},X_{ei}\mid w,\vartheta_e^\star)\bigr\}\). We provide several relevant assumptions first. 
		
	\begin{ass}[Ground truth]\label{ass:truth}
		There exists a unique \(w^\star=(\beta^\star,K^\star)\) and environment-specific
		\(\vartheta_e^\star\) generating the data according to the likelihood (\ref{eq:M})
		for all \(e\). Moreover, \(k_e(w):=\E[\ell_{ei}(w)]\ge 0\) for all \(w\),
		with equality if and only if \(w=w^\star\).
	\end{ass}
	
\begin{ass}[Prior]\label{ass:prior}
	The prior density $p(w)$ is continuous and strictly positive in a
	neighborhood of $w^\star$.
	Conditional on $w$, the nuisances $\vartheta_1,\dots,\vartheta_E$
	are a priori independent with densities that assign positive mass
	to every neighborhood of $\vartheta_e^\star$.
\end{ass}
	
\begin{ass}[Regularity]\label{ass:reg}
	The likelihood is twice continuously differentiable in $w$
	and the expectations of all first and second derivatives are finite.
	Equivalently, for all $e$ and $w$,
	$\mathbb{E}[\ell_{ei}(w)]$ and $\operatorname{Var}[\ell_{ei}(w)]$ are finite.
\end{ass}

		Let us start by establishing posterior consistency, so that credible sets shrink to neighborhoods of $w^\star$. Recall that, in general metric spaces, posterior consistency is equivalent \cite[Proposition 6.2]{ghosal2017fundamentals} to weak convergence of the posterior distribution to the Dirac measure at $w^\star$, so that the proof of Theorem \ref{thm:consistency} is about demonstrating the latter. Denote the true law by $\mathbb{P}=\mathbb{P}_1 \times \ldots \times \mathbb{P}_E$. 
		\begin{thm}[Posterior consistency]\label{thm:consistency}
			Let 	$\Pi_{n,E}(\cdot \mid D_{n,E})$ be the posterior distribution for $w$.
			Under Assumptions \ref{ass:truth}–\ref{ass:reg},  
			\[
			\Pi_{n,E}(\cdot \mid D_{n,E}) \;\rightsquigarrow\; \delta_{w^\star}
			\quad \text{in } \mathbb{P}\text{-probability as } n\to\infty,
			\]
			where $\rightsquigarrow$ denotes weak convergence of measures. 
			Equivalently, for every neighborhood \(U\) of \(w^\star\),	$\Pi_{n,E}(U^c\mid D_{n,E})\xrightarrow{\mathbb P}0.$
		\end{thm}

		As a consequence of Theorem \ref{thm:consistency}, point estimates (e.g.\ the posterior mean or mode) are consistent: they converge in probability to $(\beta^\star,K^\star)$. 
		
	The posterior distribution is itself a random object with respect to sampling: posterior consistency means that it converges to a point mass at $w^*$ both in measure-theoretic and probabilistic senses. Specifically, posterior consistency means i) weak convergence of the posterior distribution to $ \delta_{w^\star}$, ii) in probability under the true data-generating distribution $\mathbb{P}$. Weak convergence describes the limiting behavior of the posterior as a probability measure, while convergence in probability reflects the randomness induced by the data because the posterior itself is a random object. These are the two types of convergence visible in Theorem 1. 
		
		In consonance with the theoretical properties in \cite{wu}, the next result entails that the posterior
		distribution of $\|w - w^\star\|$ converges to $0$ at a particular rate which is exponential in $n$ and $E$ and improves under greater environment heterogeneity. Importantly, this is a phenomenon that the frequentist framework for causal robustness has not been able to satisfactorily account for \citep{shen2025}.
	
\begin{thm}[Posterior contraction rate]\label{thm:rate}
	Let Assumptions~\ref{ass:truth}–\ref{ass:reg} hold.
	Define
	\[
	r_{n,E}=M\sqrt{\frac{\log(nE)}{nE}}, \qquad M>0.
	\]
	Then there exist constants $C,c_1,c_2,c_{in}>0$ such that, with probability at least $1-c_1 e^{-c_2 nE}$,
	\begin{equation}\label{eq:tail}
		\Pi_{n,E}\!\bigl(\|w-w^\star\|> r_{n,E}\bigr)
		\;\le\;
		C\,\exp\!\big\{-c_2\,c_{\mathrm{in}}\,nE\,r_{n,E}^2\big\}
		\;=\;
		C\,(nE)^{-c_2 c_{\mathrm{in}} M^2}.
	\end{equation}
	Consequently, the posterior contracts around $w^\star$ at the
	parametric rate
	\[
	\|w-w^\star\|
	=O_{\Pi}\!\big((nE\,c_{\mathrm{in}})^{-1/2}\big)
	\]
	(up to a logarithmic factor), with exponential tails in $nE c_{\mathrm{in}} r^2$.
\end{thm}

The constant $c_{\mathrm{in}}$ quantifies the curvature of the Kullback–Leibler landscape around $w^\star$,
and depends on the geometric diversity of the environment means $\{\mu_e^\star\}$.
When the means are nearly linearly dependent, $\lambda_{\min}$ becomes small, making $c_{\mathrm{in}}$ small and the posterior contract more slowly.
Conversely, greater heterogeneity across environments increases $c_{\mathrm{in}}$, leading to sharper curvature and faster posterior concentration.

	\begin{remark}
		During the proof of Theorem \ref{thm:rate}, the impact of the prior pushing most probability into the neighbourhood of $w^{\star}$ has a scaling effect over the whole rate. This means that the influence of the prior is entirely outside the exponent involved in the contraction rate. 
		\end{remark}

		Finally, the well-known asymptotic properties of posterior distributions in Gaussian linear models  \citep{aos1,aos2} suggest that the posterior distribution of $w$ converges weakly to a normal distribution centered on $w^\star$. For this, we make two further assumptions.
		

		\begin{ass}[Local asymptotic normality (LAN) expansion]\label{ass:LAN}
		 For every bounded $h\in\mathbb R^{2p+1}$,
			\[
			\log\frac{p(D_{n,E}\mid w^\star+h/\sqrt N,\vartheta^\star)}{p(D_{n,E}\mid w^\star,\vartheta^\star)}
			= h^\top \Delta_N - \tfrac12 h^\top I(w^\star)h + o_{\mathbb P}(1),
			\]
			where $\Delta_N:=N^{-1/2}\sum_{e=1}^E\sum_{i=1}^{n_e}\dot\ell_{w^\star}(Z_{ei},X_{ei},Y_{ei})$. 
		\end{ass}
		
	As previously discussed in Section 2, $\|\widehat\Sigma_e-\Sigma_e\|_2=o_{\mathbb P}(n_e^{-1/2})$ for $e=1, \ldots, E$, and the contribution of the plug-in step is asymptotically negligible in the LAN expansion.
			The means $\mu_e$ are \emph{not} plugged in: they are treated as parameters with priors $\mu_e\sim\mathcal N_p(\hat\mu,\Sigma_\mu)$, where only the hyper-mean $\hat\mu$ is empirical.

		\begin{ass}\label{ass:clt}
			There exists $\delta>0$ such that 
			$\mathbb E_{w^\star,\vartheta^\star}\|\dot\ell_{w^\star}(Z,X,Y)\|^{2+\delta}<\infty$.
		\end{ass}
		
		 When the covariate dimension $p$ and the number of environments $E$ are fixed, a Bernstein-von Mises type of theorem can be established in total variation norm.

	\begin{thm}[Bernstein--von Mises]\label{thm:bvm}
		Let Assumptions \ref{ass:truth} - \ref{ass:clt} hold and  $E$ and $p$ be fixed.
		Let $\Pi_{N}(\cdot)=\Pi(\cdot\mid D_{N})$ denote the posterior for $w$. Then,
		\[
		\Bigl\|
		\Pi_{N}\!\bigl(\sqrt N\,(w-w^\star)\in\cdot\bigr)
		-
		\mathcal N\!\bigl(0,\; I(w^\star)^{-1}\bigr)
		\Bigr\|_{\mathrm{TV}}
		\;\xrightarrow[]{\mathbb P}\;0.
		\]
	\end{thm}

		\subsection{The spectrum of identifiability}

	While classical definitions frame identifiability as a binary property (a parameter is either identified or not), this paper is the first to explicitly articulate an empirical manifestation of weak identifiability as a gradual phenomenon, intimately connected to ill-posedness in inverse problems, as extensively studied recently by \cite{ghassami2025debiasedillposedregression} and \cite{neweak}. In such settings—common in instrumental variable (IV) models \citep{chib} or proximal causal inference—the underlying parameter must solve a conditional moment restriction involving a compact or nearly singular integral operator. The inversion problem becomes ill-posed and—although the model remains formally identifiable—variance inflates, confidence intervals widen, and estimation becomes unstable. In our framework, this will manifest as the prior progressively increasing its influence on the posterior as identifiability gradually blows up: a posterior contracting towards the prior reflects a violation of identification conditions, revealing the regularized nature of Bayesian posteriors. This is in contrast to frequentist estimators, which exhibit exploding variances and numerical instability in face of non-invertible or ill-conditioned design matrices. Let us make this transition from identifiability to non-identifiability visible and quantifiable, showing how Bayesian inference interpolates smoothly across this boundary. 
	
	The
	point that some parameters, even if formally identified, might have very flat
	likelihoods (and so be unidentified for practical
	purposes) is well-known \citep{gustafson}. We provide another perspective on this phenomenon  by looking at how different the posterior changed with respect to the prior as a diagnostic tool. Frequentist GI heavily relies on a key assumption: the linear independence of the environment-wise population mean vectors of the covariates. The further the identifiability condition is from being violated, the weaker the influence of the priors on $\beta$ and $K$.
	However, in settings close to linear dependence—where the frequentist method may encounter numerical issues due to matrix non-invertibility—our Bayesian formulation instead applies a controlled shrinkage towards the prior, thereby avoiding the computational instability derived from a theoretical violation of identifiability.

	In frequentist GI, when $\mu = 0$, the term $K(X_i-\mu)$ collapses to $K X_i$ making $\beta$ and $K$ unidentifiable since $\beta X_i + K X_i$ could be satisfied by infinitely many combinations of that sum to the same total. Figure \ref{violation} illustrates this phenomenon, where $\lambda$ is a parameter scaling the true mean of the one dimensional covariate, so that  $\mu \to 0$ as $\lambda \to 0$. A prior $N(0,1)$ has been placed on $\beta$. Observe how getting closer to hypothesis violation makes the prior take grip of estimation and shrink the posterior towards $0$, the prior mean. A frequentist version would just result in non-invertibility of an ill-conditioned matrix, leading to a variance explosion and possibly numerical errors.
	
	 The parameter $\mu$ enters into the model not as a nuisance to be estimated and fixed but as a random quantity. In the frequentist setting, lack of identifiability—e.g., when $\mu \to 0$—results in non-invertible design matrices and unstable estimators. Recall that in Figure \ref{violation} the posterior of $\beta$ shrinks towards its prior mean as identifiability deteriorates. While this introduces some dependence on the prior, it avoids the pathologies of the frequentist alternative. Again, this exemplifies how Bayesian methods can express the blurry nature of identifiability through the introduction of priors.  The cost, of course, is that the results will depend to some degree on the chosen prior structure.  
	
	The interpretation of causal identification within the Bayesian framework remains an open and, to some extent, philosophical question. \cite{gustafson} argue that non-identifiability manifests as posterior flatness around causal parameters. Foundational work on Bayesian inference for the LATE \citep{imbensrubin} adopts what is known as the \textit{phenomenological Bayes} approach, which views sensitivity to prior specifications as a diagnostic tool for assessing which identifying assumptions can be reasonably imposed or relaxed in IV-like settings. In \textit{Robust Statistics}, \cite{huber2009robust} write: “Instead of worrying about things not under his control, the robust Bayesian is merely concerned with inaccuracies of specification”. J. Berger in \cite{wolpert2004berger} echoes this view: “In some sense, I believe that this is the fundamentally correct paradigm for statistics—admit that the prior (and model and utility) are inaccurately specified, and find the range of implied conclusions”. 


	\section{Results}
	We begin by evaluating the proposed methodology through controlled simulated experiments of increasing complexity and sample size, contrasting its performance with standard OLS and IV estimators. We then turn to a real-world case study analyzing the effect of lifestyle factors on BMI. All artificial experiments and algorithmic implementations described in this work can be reproduced using the publicly available code at \url{https://github.com/meixide/deconfounded_posterior}.
	\subsection{Simulated experiments}\label{sec:sim}
	Let us first demonstrate, through a simple one-dimensional ($p=1$), no intercept and single training source ($E=1$) setup, the advantages of our proposed method in comparison to OLS and IV. The second part is more computationally intensive, numerically comparing the empirical coverage for predictions generated through our approach and OLS across multiple simulation runs.
	\subsubsection{Single source example}
	
\begin{figure}[t!]
  \centering
  
  \begin{subfigure}[b]{0.48\linewidth}
    \centering
    \includegraphics[width=0.8\linewidth]{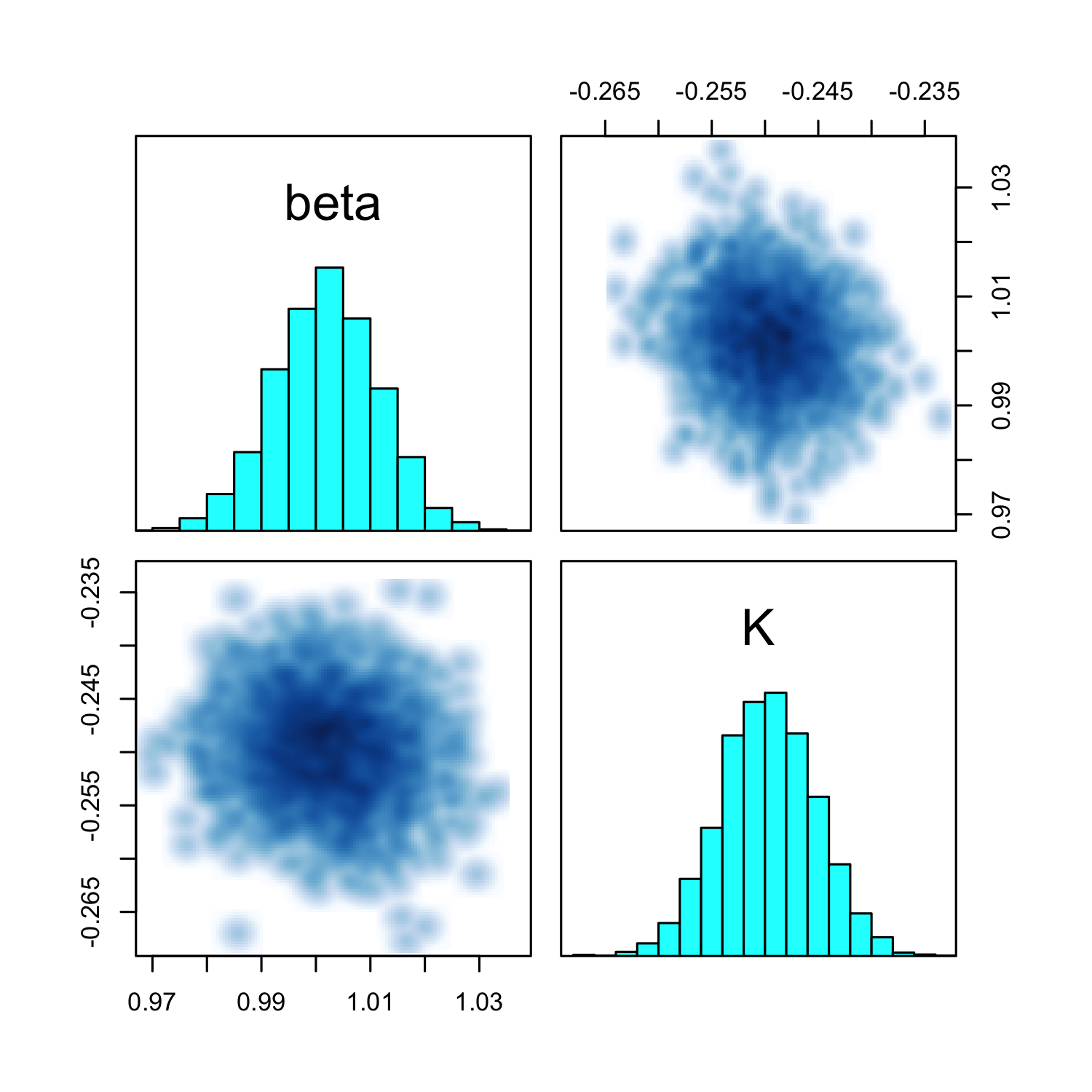}
    \caption{}
    \label{fig:stan}
  \end{subfigure}
  \hfill
  \begin{subfigure}[b]{0.48\linewidth}
    \centering
    \includegraphics[width=0.8\linewidth]{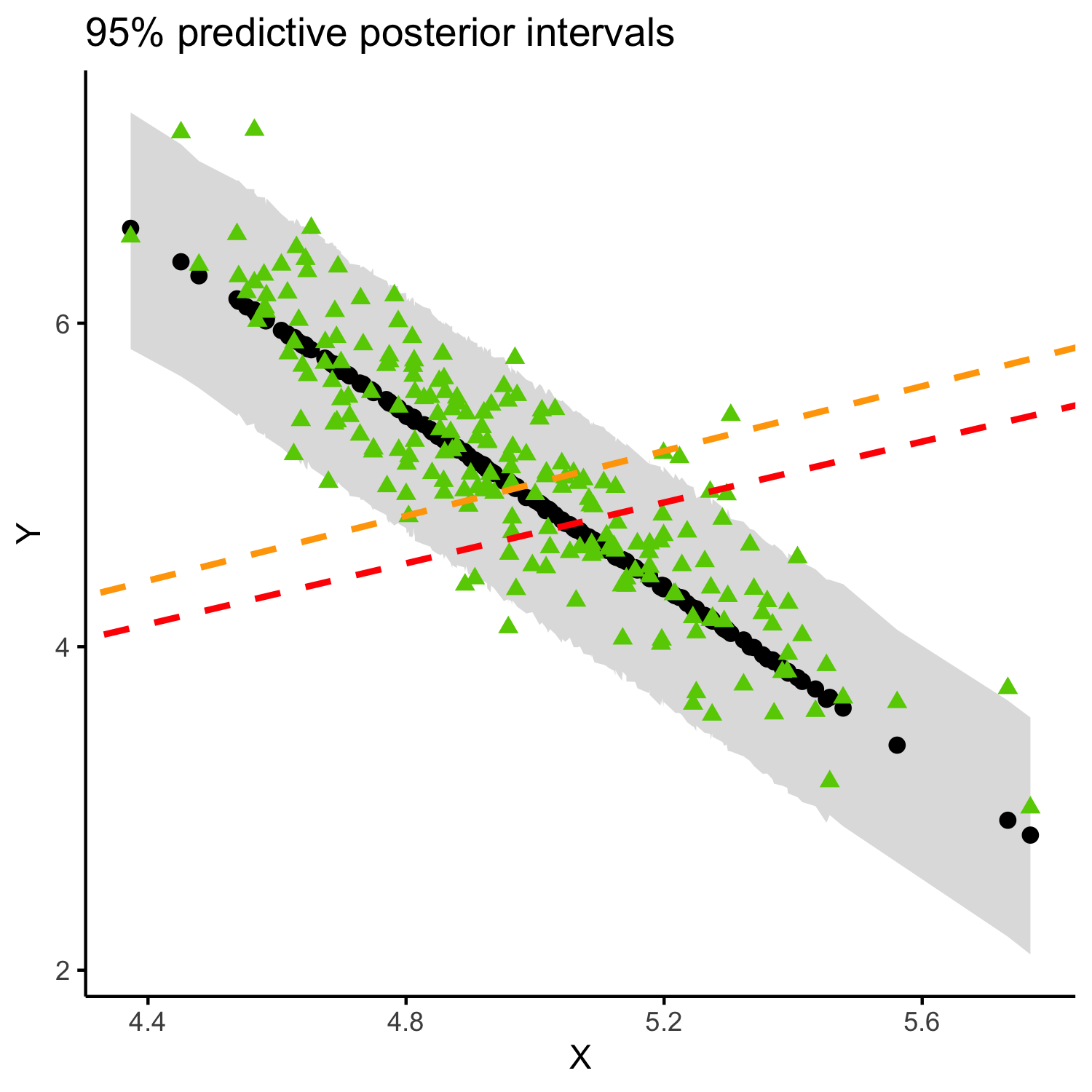}
    \caption{}
    \label{fig:pred}
  \end{subfigure}
  
  \caption{Plots of simulation results from Section 6.1.1. a) posterior distributions of $\beta$ and $K$ for the single source example. b) our proposal's predictive credible intervals in grey and the posterior mean in black. Green: unseen response sample. Red and orange: best OLS and IV predictions. Empirical coverage, 0.96.}
  \label{fig:combined}
\end{figure}

	We generate one training data source with the following specifications. The number $n_1$ of observations is
     $500$. 
	A confounder $H \sim N(0, \sigma_H^2)$ is generated 
     with $\sigma_H = 0.5$. The predictor $X$ depends linearly on $H$ with $X = 0.5H + \varepsilon_X$, and  $\varepsilon_X \sim N(0, 0.1^2)$. 
	The outcome $Y$ is influenced by both $X$ and $H$:
	$Y = X - 2H + \varepsilon'_Y =: X + \varepsilon_Y, \quad \varepsilon'_Y \sim N(0, 0.01^2)$. In this case, $K^*= \operatorname{Cov}(- 2H + \varepsilon'_Y, 0.5H + \varepsilon_X) = -2 \cdot 0.5 \operatorname{Var} H = -0.25$. However, the user ignores $H$, misspecifying the model by ommiting it. All three estimators in the present comparison explicitly exclude an intercept. This choice is deliberate: including an intercept would require at least two environments for GI identifiability, compromising the transparency of the illustration (the primary objective of this experiment). Summary statistics from the OLS estimation for the slope yield $\hat{\beta} = 0.936$ 
    with standard error $0.015$, far from the actual value $\beta_* = 1$. In contrast, Figure \ref{fig:stan} shows that the joint and marginal posterior distributions for $\beta$ and $K$, obtained with \texttt{Rstan}, are correctly centered at $1$ and $-0.25$, respectively.

	Next, a smaller testing dataset of size $n_0 = 200$ is generated with shift $3$ applied to $X$ with respect to the training environment distribution. Figure \ref{fig:pred} suggests that both OLS (red) and IV (orange) are useless for making predictions. Notice how the 2.5\% and 97.5\% quantiles of our posterior predictive distribution, in grey, accurately capture the true unseen data (green triangles), where only the $X$ coordinate is observed in a non-simulated setting.
		
Note that the GI estimator predictive posterior is not equivalent to standard IV estimation using the domain indicator as an IV even though the posterior of $\beta$ is actually centered on the IV estimator using the domain indicator as an instrument. The key insight is that $K$, absorbing the effects of the hidden confounders, is also identifiable as a function of $\beta$, being essential for completing the conditional expectation of $Y$ given $X$ in the unseen environment. Using an IV under proper assumptions suffices to identify the causal posterior mean of $\beta$ but is not useful for prediction purposes. As Figure \ref{fig:pred} shows, although IV (orange line) is the causal truth it renders completely wrong for making predictions, being far apart from predicting the green points. We emphasize that the comparison in Figure \ref{fig:pred} is fair, as all three estimators were deliberately specified without an intercept in this experiment.
	\subsubsection{Multiple sources example}
	
	Data generation is now designed to mimic scenarios where observed covariates \(X\) and response \(Y\) are simultaneously influenced by unobserved confounders \(H\), and the relationships vary across different environments. For each training environment \(e \in \{1, \dots, E\}\), we generate \(\lceil \frac{n}{E} \rceil \) observations of $(X,Y)$. The $q$ hidden confounders \(H \in \mathbb{R}^{n \times q}\) are drawn from a standard normal \(H_{i, .} \sim N(0, I_q)\). The environment-specific covariate component  \(V \in \mathbb{R}^{n \times p}\) is  generated from a multivariate normal distribution with mean vector specific to the environment \(z\) and a covariance matrix \(\Sigma_V\), so that $V_{i, .} \sim N(\mu_e, \Sigma_V)$,  $\Sigma_V = 0.5 \cdot ({1}_p {1}_p^\top) + 0.5 \cdot I_p$. The covariates \(X \in \mathbb{R}^{n \times p}\) are generated as a linear combination of the hidden confounders and the environment-specific part $X_{i, .} = V_{i, .} + \Psi^\top H_{i, .} $, where \(\Psi \in \mathbb{R}^{q \times p}\) is a coefficient matrix encoding the influence of the hidden confounders on the covariates. \(Y\) is generated as a linear combination of the observed covariates \(X\), the hidden confounders \(H\), and noise: $Y_i = \beta ^T X_{i, .}  +\phi ^T H_{i, .}  + \varepsilon'_{Y,i}$, where \(\beta \in \mathbb{R}^p\) is the 
      coefficient vector, \(\phi \in \mathbb{R}^q\) encodes the effect of the hidden confounders on the response, and \(\varepsilon'_{Y,i} \sim N(0, 1)\) is independent noise.  The mean vectors for the environment-specific covariates 
        are systematically varied across environments, with $
\mu_j = \frac{2j}{p} - 1 + u_j, \quad \text{and } u_j \sim \text{Unif}(-1,1), \quad j = 1, \dots, p,
$ drawn once per dataset (deterministic ground truth). This process is repeated for \(E = p + 1\) environments to enable identifiability. A separate test dataset is generated using the fixed mean vector
$\mu^0_{j} =  \frac{2j}{p} +2 \cdot U, \quad \text{with } U \sim \text{Unif}([-1,1]^p)$,
and a perturbed covariance matrix
$
\Sigma^{0} = \Sigma_V + 0.5 \cdot I_p.
$

\begin{figure}[htbp]
	\centering
	\begin{minipage}{0.45\textwidth}
		\centering
		\begin{tabular}{l|ccc}
			\hline
			$n,p$ & 2 & 5 & 10 \\ \hline
			200  & .88/.96   & .85/.95   & .87/.90      \\
			500  & .91/.95   & .88/.93   & .83/.94      \\
			1000 & .89/.95   & .88/.95   & .85/.94      \\
			2000 & .90/.95   & .83/.94   & .80/.95      \\
			\hline
		\end{tabular}
		\caption{Average empirical coverage for OLS vs our approach for the setup in 3.1.2 depending on sample size $n$ and 
			number $p$ of parameters with $\alpha=0.05$}
		\label{table:AEC}
	\end{minipage}\hfill
	\begin{minipage}{0.5\textwidth}
		\centering
		\includegraphics[angle=90,width=\linewidth]{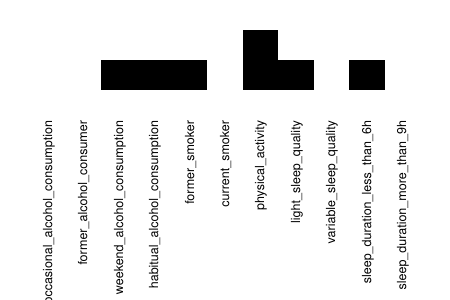}
		\caption{Black squares indicate selected lifestyle factors. Top row corresponds to our estimator; while bottom row represents standard OLS rule with \textit{p}-values $<0.05$.}
		\label{fig:quiron}
	\end{minipage}
\end{figure}

	We utilized 8 cores, running 3 simulations per core, (24 runs). For each run, we obtained a posterior sample of size $4$ chains $\times$ $1000$ iterations $= 4000$ for every test observation of the covariates. For each of these samples, we computed the 2.5\% and 97.5\% quantiles and assessed whether the true response fell within this interval. We then calculated the proportion of predictive intervals containing the ground truth across all 24 runs, which, ideally, should be close to 0.95. As a baseline, we computed standard large-sample normal approximation confidence intervals for OLS estimates. Table \ref{table:AEC} presents the average empirical coverage across simulation runs for various sample sizes and dimensions, comparing 
     OLS (left) and ours (right), outperforming OLS
      in all cases. Even more, our coverage remains largely stable as $p$ increases, whereas OLS exhibits a clear deterioration.

	

	
	\subsection{Case study: analysis of lifestyle factors affecting BMI} \label{sec:quiron}
	
	Consider analysing the causal relationship between lifestyle factors including alcohol consumption, smoking habits, sleep quality, and sleep duration ($X$), and BMI ($Y$) using a comprehensive proprietary dataset with individuals from various Spanish provinces. The dataset encompasses as well numerous potential confounding variables, including sex, cholesterol and glucose levels treated in our study, for illustration purposes, as hidden variables affecting both $X$ and $Y$ simultaneously. The province from which an observation originates serves as the dataset indicator $Z$.
	
	Our framework applied to this case relies on two assumptions: the confounding structure remains consistent across different provinces (akin to the inner-product invariance assumption), so that the distribution of hidden variables and their relationship with our variables of interest is invariant geographically; and the absence of direct causal paths from location to BMI, so that any influence of $Z$ on $Y$ must be mediated through $X$. This aligns with the structural equation model (1), with province serving as IV \citep{angrist1996}.

	The use of geographical location as an instrument finds precedent in the medical literature, with notable examples including \citep{beck2003, leigh2004, mcclellan1994}. In our analysis, Spanish coastal peninsular provinces are considered the training environments and evaluate BMI predictions in Madrid, in the center of the country. While we benefit from access to true BMI values in the test domain for validation, we acknowledge that such complete information is typically unavailable in real-world applications, where health researchers often just have access to covariate samples in test domains. Our analysis achieved a remarkable empirical coverage of $0.95$, with physical activity emerging as the sole causal variable using the selection criterion in Section \ref{sec:sel}. 
	
    This stands in contrast to traditional OLS analysis, which identified multiple significant correlations at the $\alpha=0.05$ level. For example, consider the apparently positive effect of the variable \texttt{former\_smoker}, which clearly illustrates the strength of our methodology: quitting smoking is correlated with weight gain, although it does not causally influence BMI. While OLS indicates a significant correlation between smoking cessation and increased BMI, potentially reflecting the well-documented phenomenon of post-cessation weight gain due to anxiety-induced eating \citep{Aubine4439}, our analysis suggests this relationship as correlative rather than causal. This distinction highlights our methodology's capability to discriminate between genuine causal effects and mere correlations, while preserving predictive performance.
	
	

	\section{Discussion}

We have introduced a new framework that delivers calibrated probabilistic predictions and credible causal discovery simultaneously, even when distribution shifts and hidden confounders undermine conventional methods. Unlike classical IV or minimax-robust approaches—which either sacrifice unbiasedness through regularization or offer no uncertainty quantification—we introduce a hierarchical prior to fuse information from multiple training environments and unlabeled target data, yielding fully Bayesian posteriors for both causal parameters and future outcomes. These priors endow the model with rigorous asymptotic guarantees, turning the number of distinct datasets into a meaningful measure of precision. Moreover, our approach enables the first practical hypothesis test for causal parenthood under hidden confounding. Extensive simulations and a real medical case study suggest that our method maintains almost nominal coverage from low to moderate dimensions, while OLS and classical IV fail to generalize.

 We acknowledge that our Bayesian formulation assumes Gaussianity for the covariates in the likelihood model. This assumption, however, should not be interpreted as an essential requirement of our framework. Rather, it is a modeling convenience much like the standard Gaussian assumption on error terms in classical Bayesian linear regression. An important avenue for future work is the integration of the martingale posterior distribution framework recently proposed by \cite{mart_post}, enabling the construction of posterior distributions without the traditional need of specifying likelihood and priors. This perspective reinterprets Bayesian uncertainty as arising from the distribution of unobserved data, rather than from prior uncertainty on parameters by focusing on predictive distributions over missing observations and leveraging martingale theory. 
 
 The adversarial machine learning (AML) arena \citep{cina2023wild} considers situations in which an adversary modifies the operational data distribution in pursue of some objective, thereby inducing a distribution shift. Bayesian approaches \citep{rios2023} seem natural to model the available partial information about how the attacker might have perturbed such a distribution. Moreover, as attackers interact with an ML system, they learn how to modify their attacks inducing further shifts. A natural extension of this work involves modeling adversarial attacks in ML scenarios where the attacker manipulates only the covariates with the goal of steering the posterior predictive mean towards a desired target value. If the attacker’s target is modeled probabilistically, this uncertainty induces a distribution over the perturbations (and, consequently, over the manipulated inputs) that could then be used to update prior beliefs about the regression coefficients and other model parameters. A promising next step would be to unify this framework with ours and investigate how such adversarial perturbations can inform or refine prior distributions in that setting.

	Another promising direction consists of relaxing the assumption that the confounder's distribution and their influence on observed variables remains constant across environments. Similar to our approach, where environment means are naturally sampled from a prior rather than relying on frequentist estimators, thereby easing strict frequentist identification conditions, each environment could be assigned a distinct $K^z_*$, all shrunk through a prior. More broadly, further investigation into relaxing identifiability conditions is needed and would have a significant impact on the field of causal representation learning, which currently relies on the assumption that the number of environments must at least equal the number of latent nodes \citep{jin}.
	
	Additional theoretical questions remain open for future investigation. First, when generalizing to a new domain we
     should check whether GI achieves a marginal PAC guarantee \citep{zhang2024}. 
     Specifically, for $(X_{n+1}, Y_{n+1}) \sim \mathbb{P}_0$, 
      we would like to establish whether $\mathbb{P}\left\{Y_{n+1} \notin \hat{C}_n\left(X_{n+1}\right) \mid D_N\right\} \leq \alpha+o_{\mathbb{P}}(1)$ with high probability
	or even a stronger guarantee conditional on $X_{n+1}$, possibly using the martingale posterior framework \citep{mart_post}. This approach could be particularly powerful as it generalizes traditional Bayesian posteriors while eliminating the need for explicit likelihood-prior construction.
	
%
%
%
%
%
%
%


\appendix
	\section*{Supplementary A: theorem proofs}
\subsection*{Proof of Theorem \ref{thm:consistency}}

Let
\[
S_{n,E}(w)=\sum_{e=1}^E\sum_{i=1}^{n_e} \ell_{ei}(w),
\qquad
\ell_{ei}(w)
=
\log p_e(Y_{ei}\mid X_{ei},w^\star,\vartheta_e^\star)
-
\log p_e(Y_{ei}\mid X_{ei},w,\vartheta_e^\star),
\]
and define
\[
k_e(w):=\E[\ell_{ei}(w)],\qquad
k(w):=\frac{1}{E}\sum_{e=1}^E k_e(w).
\]
Under the linear–Gaussian model and identifiability in Assumption~\ref{ass:truth}, $k(w)\ge 0$ with $k(w)=0$ iff $w=w^\star$, and in fact
$k(w)$ is quadratic in $w-w^\star$ with a positive–definite curvature at $w^\star$. By Assumption~\ref{ass:reg} (finite second moments and smoothness)
and standard parametric LLN, for every compact $\mathcal{K}\subset\mathbb{R}^{2p+1}$,
\begin{equation}\label{eq:uniflln}
	\sup_{w\in\K}\Bigg|
	\frac{1}{N}\Big(S_{n,E}(w)-S_{n,E}(w^\star)\Big)-k(w)
	\Bigg|\xrightarrow{\mathbb P}0,
	\qquad N:=\sum_{e=1}^E n_e.
\end{equation}

The posterior measure of a Borel set $A\subset\mathbb{R}^{2p+1}$ is
\[
\Pi_{N}(A\mid D_N)
=\frac{\int_A p(w)\,e^{-S_{n,E}(w)}\,dw}{\int_{\mathbb{R}^{2p+1}} p(w)\,e^{-S_{n,E}(w)}\,dw}.
\]
To prove $\Pi_{N}(\cdot\mid D_N) \rightsquigarrow\delta_{w^\star}$ in $\mathbb P$-probability, it is
enough to show that for every neighborhood $U$ of $w^\star$,
\begin{equation}\label{eq:target}
	\Pi_{N}(U^c\mid D_N)\xrightarrow{\mathbb P}0.
\end{equation}

By continuity of $k$ and $k(w)=0$ only at $w^\star$ , we have
\[
 d:=\inf_{w\in U^c} k(w)>0.
\]
Fix $\varepsilon\in(0, d)$. Applying the uniform LLN \eqref{eq:uniflln} on
$\K=U^c\cap B_R$ and letting $R\to\infty$, we get that with probability
tending to $1$,
\begin{equation}\label{eq:sep}
	S_{n,E}(w)-S_{n,E}(w^\star)\ \ge\ N\,( d-\varepsilon)
	\qquad\forall\,w\in U^c.
\end{equation}

We derive an upper bound for the numerator. On the event \eqref{eq:sep},
\[
\int_{U^c} p(w)\,e^{-S_{n,E}(w)}\,dw
\ \le\
e^{-N( d-\varepsilon)}\int_{U^c} p(w)\,e^{-S_{n,E}(w^\star)}\,dw
\ \le\ e^{-N( d-\varepsilon)}\,e^{-S_{n,E}(w^\star)}.
\]

We lower bound the denominator now. By the prior positivity in Assumption~\ref{ass:prior}, there exist $r_0>0$ and $P_{\min}>0$
such that $p(w)\ge P_{\min}$ for all $w\in B_{r_0}(w^\star)$. Using the uniform LLN
\eqref{eq:uniflln} on $\K=B_{r_0}(w^\star)$, possibly shrinking $r_0$ and with the same
$\varepsilon>0$, with probability tending to $1$,
\[
\sup_{w\in B_{r_0}} \{S_{n,E}(w)-S_{n,E}(w^\star)\}\ \le\ N\,\varepsilon,
\]
and hence
\[
\int_{\mathbb{R}^{2p+1}} p(w)\,e^{-S_{n,E}(w)}\,dw
\ \ge\ \int_{B_{r_0}} p(w)\,e^{-S_{n,E}(w)}\,dw
\ \ge\ P_{\min}\,|B_{r_0}|\,e^{-N\varepsilon}\,e^{-S_{n,E}(w^\star)}.
\]

Combining the two bounds, with probability tending to $1$,
\[
\Pi_{N}(U^c\mid D_N)
\ \le\
\frac{e^{-N( d-\varepsilon)}}{P_{\min}\,|B_{r_0}|\,e^{-N\varepsilon}}
\ =\
C\,e^{-N( d-2\varepsilon)},
\qquad C:=(P_{\min}|B_{r_0}|)^{-1}.
\]
Since $\varepsilon\in(0, d/2)$ is arbitrary, the right-hand side converges to $0$
in $\mathbb P$-probability, proving \eqref{eq:target}. Therefore,
\[
\Pi_{N}(\cdot\mid D_N)\  \rightsquigarrow\ \delta_{w^\star}
\quad\text{in }\mathbb P\text{-probability as }N\to\infty.
\]

								\subsection*{Proof of Theorem \ref{thm:rate}}
								
								Let
								\[
								S_{n,E}(w)=\sum_{e=1}^{E}\sum_{i=1}^{n_e}\ell_{ei}(w),\qquad
								\ell_{ei}(w)=\log\!\frac{p_e(Y_{ei}\mid X_{ei},w^\star,\vartheta_e^\star)}
								{p_e(Y_{ei}\mid X_{ei},w,\vartheta_e^\star)},
								\]
								and define
								\[
								k_e(w):=\E[\ell_{ei}(w)],\qquad
								k(w):=\frac{1}{E}\sum_{e=1}^E k_e(w),\qquad
								N:=\sum_{e=1}^E n_e.
								\]
							By the linear–Gaussian model and identifiability (Assumption~\ref{ass:truth}),
							$k(w)\ge 0$ with $k(w)=0$ iff $w=w^\star$. Moreover, there exists $r_0,c_{in}>0$ such that
							\begin{equation}\label{eq:local-quad}
								k(w)\;\ge\; c_{\mathrm{in}}\|w-w^\star\|^2
								\qquad\text{whenever }\|w-w^\star\|\le r_0 .
							\end{equation}
							To see this, set $u:=w-w^\star$ and define the environment–specific feature vector
							\[
							\phi_e(X):=
							\begin{bmatrix}
								1\\[2pt] X\\[2pt] X-\mu_e^\star
							\end{bmatrix}
							\in\mathbb R^{2p+1},
							\qquad
							\Delta m_e(X;w,w^\star)
							:= \phi_e(X)^\top u .
							\]
							With Gaussian observation noise of variance $\sigma_Y^{2\star}$, the per–environment
							Kullback–Leibler gap equals the scaled mean–squared prediction error:
							\[
							k_e(w)
							=\frac{1}{2\sigma_Y^{2\star}}\,
							\mathbb E_e\!\left[(\Delta m_e(X;w,w^\star))^2\right]
							=\frac{1}{2\sigma_Y^{2\star}}\,u^\top S_e\,u,
							\qquad
							S_e := \mathbb E_e[\phi_e(X)\phi_e(X)^\top].
							\]
							Averaging over environments,
							\[
							k(w)
							=\frac{1}{E}\sum_{e=1}^E k_e(w)
							=\frac{1}{2\sigma_Y^{2\star}}\,u^\top S\,u,
							\qquad
							S := \frac{1}{E}\sum_{e=1}^E S_e .
							\]
							By Rayleigh–Ritz for the symmetric positive definite matrix $S$, we have
							\[
							u^\top S u \;\ge\; \lambda_{\min}(S)\,\|u\|^2.
							\]
							Hence
							\[
							k(w)\;\ge\;\frac{1}{2\sigma_Y^{2\star}}\,\lambda_{\min}(S)\,\|w-w^\star\|^2,
							\]
							so the curvature constant can be taken as
							\[
						c_{\mathrm{in}}
								=\frac{1}{2\sigma_Y^{2\star}}\,\lambda_{\min}(S),
								\qquad
								S=\frac{1}{E}\sum_{e=1}^E \mathbb E_e[\phi_e(X)\phi_e(X)^\top].
							\]
							(If desired, \eqref{eq:local-quad} can be stated globally since the bound above holds for all $w$.)
								If the noise variances differ across environments,
								the same argument gives
								\[
								c_{\mathrm{in}}
								=\frac{1}{2}\,
								\lambda_{\min}\!\Bigl(
								\frac{1}{E}\sum_{e=1}^E
								\sigma_{Y,e}^{-2}\,
								\E_e[\phi_e(X)\phi_e(X)^\top]
								\Bigr).
								\]
								Assumption~\ref{ass:reg} (finite second moments and smoothness) implies a
								\emph{uniform} law of large numbers on compacta for the parametric family:
								for every compact $\mathcal{K}\subset\mathbb{R}^{2p+1}$,
								\begin{equation}\label{eq:ULLN}
									\sup_{w\in\K}\Bigg|
									\frac{1}{N}\Big(S_{n,E}(w)-S_{n,E}(w^\star)\Big)-k(w)
									\Bigg|\xrightarrow{\;\mathbb P\;}0 .
								\end{equation}
								
								Fix $0<r\le r_0$ and write $U:=B_r(w^\star)$, $U^c=\{w:\|w-w^\star\|>r\}$.
								By continuity of $k$ and uniqueness of its minimizer at $w^\star$, the global
								separation outside $U$ is strictly positive:
								\[
								k_{\min}(r):=\inf_{\|w-w^\star\|\ge r} k(w)\ >\ 0.
								\]
								
							We provide an upper bound for the numerator. Let $\varepsilon\in(0,k_{\min}(r))$ and apply \eqref{eq:ULLN} on $\K=U^c\cap B_R$,
								then let $R\to\infty$. With probability tending to $1$,
								\[
								\inf_{w\in U^c}\big\{S_{n,E}(w)-S_{n,E}(w^\star)\big\}
								\ \ge\ N\{k_{\min}(r)-\varepsilon\},
								\]
								and therefore
								\[
								\int_{U^c} p(w)\,e^{-S_{n,E}(w)}\,dw
								\ \le\ e^{-N(k_{\min}(r)-\varepsilon)}\int_{U^c} p(w)\,e^{-S_{n,E}(w^\star)}\,dw
								\ \le\ e^{-N(k_{\min}(r)-\varepsilon)}\,e^{-S_{n,E}(w^\star)}.
								\]
								
							Next, we derive a lower bound for the denominator.  By Assumption~\ref{ass:prior}, there exist $r_1\in(0,r_0]$ and $P_{\min}>0$
								such that $p(w)\ge P_{\min}$ on $B_{r_1}(w^\star)$. Using \eqref{eq:ULLN}
								on $\K=B_{r_1}(w^\star)$ (possibly shrinking $r_1$) and the same $\varepsilon>0$,
								with probability tending to $1$,
								\[
								\sup_{w\in B_{r_1}}\{S_{n,E}(w)-S_{n,E}(w^\star)\}\ \le\ N\varepsilon,
								\]
								hence
								\[
								\int_{\mathbb{R}^{2p+1}} p(w)\,e^{-S_{n,E}(w)}\,dw
								\ \ge\ \int_{B_{r_1}} p(w)\,e^{-S_{n,E}(w)}\,dw
								\ \ge\ P_{\min}\,|B_{r_1}|\,e^{-N\varepsilon}\,e^{-S_{n,E}(w^\star)}.
								\]
								
								Combining both bounds, with probability tending to $1$,
								\[
								\Pi_{N}(U^c\mid D_N)
								=\frac{\int_{U^c} p(w)\,e^{-S_{n,E}(w)}\,dw}{\int p(w)\,e^{-S_{n,E}(w)}\,dw}
								\ \le\ 
								\frac{e^{-N(k_{\min}(r)-\varepsilon)}}{P_{\min}\,|B_{r_1}|\,e^{-N\varepsilon}}
								\ =\ C\,\exp\{-N\,(k_{\min}(r)-2\varepsilon)\},
								\]
								with $C:=(P_{\min}|B_{r_1}|)^{-1}$.

								By the local quadratic lower bound \eqref{eq:local-quad}, for $r\le r_0$,
								$k_{\min}(r)\ge c_{\mathrm{in}} r^2$. Taking $\varepsilon=\tfrac14 c_{\mathrm{in}}r^2$ gives
								\[
								\Pi_{N}\!\big(\|w-w^\star\|> r\big)
								\ \le\ C\,\exp\!\Big\{-\tfrac12 c_{\mathrm{in}}\,N\,r^2\Big\}
								\qquad\text{with probability converging to }1
								\]
								Finally, choose
								\(
								r_{n,E}=M\sqrt{\frac{\log N}{N}}
								\)
								to obtain
								\[
								\Pi_{N}\!\big(\|w-w^\star\|> r_{n,E}\big)
								\ \le\ C\,N^{-\frac12 c_{\mathrm{in}} M^2}
								\qquad\text{with probability converging to }1,
								\]
								which is the claimed parametric contraction (up to a $\sqrt{\log N}$ factor).
				\subsection*{Proof of Theorem \ref{thm:bvm}}
				
We follow standard BvM steps (\cite{vandervaart1998asymptotic} chapter 10; \cite{klejn}). Recall $N=\sum_{e=1}^En_e$ and denote by $\Pi_{N}(\cdot)=\Pi(\cdot\mid D_{N})$ the posterior for $w$.
					Let $h=\sqrt N\,(w-w^\star)\in\mathbb R^{2p+1}$ and denote by
					\[
					\Pi_N^h(\cdot):=\Pi_{N}\bigl(\sqrt N\,(w-w^\star)\in\cdot\bigr)
					\]
					the posterior distribution of $h$. We show that $\Pi_N^h$ converges in total variation to
					$\mathcal N(I(w^\star)^{-1}\Delta_N,\,I(w^\star)^{-1})$.

					By Assumption~\ref{ass:LAN}, we have the following LAN expansion
					uniformly for $\|h\|\le M$ (fixed $M$),
					\begin{equation}\label{eq:LAN}
						\log\frac{p(D_{N}\mid w^\star+h/\sqrt N,\vartheta^\star)}
						{p(D_{N}\mid w^\star,\vartheta^\star)}
						= h^\top \Delta_N - \tfrac12 h^\top I(w^\star) h + r_N(h),
						\quad\text{with } \sup_{\|h\|\le M}|r_N(h)|=o_{\mathbb P}(1).
					\end{equation}
					The plug--in of $\widehat\Sigma_e$ (and the full-Bayes treatment of $\mu_e$) only contributes $o_{\mathbb P}(1)$ to \eqref{eq:LAN}; hence the same expansion holds for the marginal likelihood in $w$.

					By Assumption~\ref{ass:prior}(i) and continuity of $\pi(w)$ at $w^\star$, we have the following prior flatness property
					\[
					\pi(w^\star+h/\sqrt N)=\pi(w^\star)+o(1)
					\qquad\text{uniformly for }\|h\|\le M.
					\]

				The local likelihood ratio is
					\[
					\pi(w^\star+h/\sqrt N)\,p(D_{N}\mid w^\star+h/\sqrt N,\vartheta^\star)
					= \pi(w^\star)\,p(D_{N}\mid w^\star,\vartheta^\star)\,
					\exp\Bigl\{h^\top\Delta_N-\tfrac12 h^\top I(w^\star)h + o_{\mathbb P}(1)\Bigr\},
					\]
					uniformly on $\|h\|\le M$.

					By Theorem 2, for every $M\to\infty$,
					\[
					\Pi_N^h(\|h\|>M)=o_{\mathbb P}(1).
					\]
					Hence we can restrict attention to $\|h\|\le M$ with probability $\to1$, and the $o_{\mathbb P}(1)$ remainder is uniform there.

					Now, let 
					\[
					\phi_N(h):=\exp\Bigl\{h^\top\Delta_N-\tfrac12 h^\top I(w^\star)h\Bigr\},\qquad
					\phi(h):=\exp\Bigl\{-\tfrac12 h^\top I(w^\star)h\Bigr\}.
					\]
					The posterior density (w.r.t.\ Lebesgue on $h$) on $\|h\|\le M$ is, up to a $1+o_{\mathbb P}(1)$ factor,
					\[
					\frac{\phi_N(h)}{\int_{\|u\|\le M}\phi_N(u)\,du}.
					\]
					Since $\Delta_N  \rightsquigarrow \mathcal N(0,I(w^\star))$ by Assumption~\ref{ass:clt}, and $\phi_N$ is the likelihood of a normal shift experiment, standard arguments (e.g.\ Theorem 10.1 in \cite{vandervaart1998asymptotic}) imply
					\[
					\sup_{A} \Bigl|
					\int_A \frac{\phi_N(h)}{\int\phi_N}\,dh
					-
					\int_A \varphi_{I(w^\star)^{-1}\Delta_N,\ I(w^\star)^{-1}}(h)\,dh
					\Bigr| = o_{\mathbb P}(1),
					\]
					where $\varphi_{m,V}$ is the density of $\mathcal N(m,V)$. This yields
					\[
					\bigl\|\Pi_N^h - \mathcal N(I(w^\star)^{-1}\Delta_N,\ I(w^\star)^{-1})\bigr\|_{\mathrm{TV}}
					=o_{\mathbb P}(1).
					\]

					Because of Assumption \ref{ass:clt}, $\Delta_N \rightsquigarrow \mathcal N(0,I(w^\star))$ by the Lindeberg–Feller CLT. Therefore, Slutsky’s theorem gives
					\[
					\mathcal N\!\bigl(I(w^\star)^{-1}\Delta_N,\ I(w^\star)^{-1}\bigr)
					\;\xrightarrow[\mathrm{TV}]{\mathbb P}\;
					\mathcal N\!\bigl(0,\ I(w^\star)^{-1}\bigr).
					\]
					The triangle inequality for total variation then implies the centered version in the statement of the Theorem. 
					 \hfill $\square$

\section{Supplement B: computational complements}
\subsection*{Single predictor, single test environment}\label{sec:onedim}

This section considers first a single-predictor, single-training source setup, providing 
a Bayesian formulation of the least-squares problem in \cite{meixide2025unsuperviseddomainadaptationhidden} as well as expressions 
for the required conditional posterior distributions,
serving to
derive theoretical insights while preserving the notation streamlined.
We then consider the multiple predictor, multiple test environment setup, formulating the model, discussing its computational implementation, and illustrating relevant applications.

Let us provide first a Bayesian analysis of the basic setup (1) to obtain the posterior predictive distributions required to issue predictions for $Z=0$.
We have available  training  
${D}_1 $ 
and testing  
${D}_0$ 
data, 
with our objective being to predict the (missing) labels
$\{ Y_{0,1},...,Y_{0,n_0} \}$ in this testing environment.
In the training environment, let us formulate the likelihood
$$\begin{aligned}Y_i= \gamma X_i +K ( X_i - \mu) + \varepsilon'_{Y,i}, \quad &\varepsilon'_{Y,i}\sim N (0, \sigma_Y ^2),  \\ 
	X_i \sim N (\mu , \sigma ^2 _X), \quad &i =1,\ldots ,n 
\end{aligned}$$
and complete it with standard normal-inverse gamma priors 
\citep{french2000statistical} from one-dimensional Bayesian linear regression models: 
\[\begin{aligned} \mu | \sigma^2 _\mu ,\tau^2_\mu \sim N(\hat \mu , \tau_{\mu}^2 \sigma_
	\mu^2),\quad &\gamma, K | \sigma^2_Y, \tau^2_ Y  \sim N(0, \tau_Y ^2\sigma_Y^2  ), \\  
	\tau_{\mu} ^2 \sim \operatorname{Inv-Gamma} (a_{\tau_{\mu} },b_{\tau_{\mu} }),\quad  &  	\sigma_{\mu} ^2 \sim \operatorname{Inv-Gamma} (a_\mu, b_\mu),  \\
	\tau _Y ^2 \sim \operatorname{Inv-Gamma} (a_{\tau_Y}, b_{\tau_Y }),	\quad &	\sigma_Y ^2 \sim \operatorname{Inv-Gamma} (a_Y, b_Y), \\ &\sigma ^2_X \sim \operatorname{Inv-Gamma} (a_X,b_X),
	&
\end{aligned}\]

\noindent with $\hat \mu$ the sample average of the covariate. 

Using standard arguments, the marginal posteriors for the variance parameters $\tau_Y^2, \tau_X^2, \sigma_Y^2, \sigma_X^2, \sigma_\mu^2$ follow inverse-gamma distributions. 
Given $\mu, \tau^2_Y, \sigma^2_Y$, the joint marginal posterior of $\gamma$ and $K$ up to an additive constant that does not depend on the parameters is 
\begin{align*}	\log p(\gamma ,K \mid \mu,\tau_Y^{2},\sigma_Y^{2},D_1)
	&\;\propto\;
	-\frac{1}{2\sigma_Y^{2}}
	\Bigl[
	\Bigl(\textstyle\sum_{i=1}^{n}X_i^{2}+\frac{1}{\tau_Y^{2}}\Bigr)\gamma^{2}
	+
	\Bigl(\textstyle\sum_{i=1}^{n}(X_i-\mu)^{2}+\frac{1}{\tau_Y^{2}}\Bigr)K^{2} \\
	&+
	2\gamma K\sum_{i=1}^{n}X_i(X_i-\mu)
	\Bigr] \\
	&
	+\,\frac{\gamma}{\sigma_Y^{2}}\sum_{i=1}^{n}Y_iX_i
	+\frac{K}{\sigma_Y^{2}}\sum_{i=1}^{n}Y_i\bigl(X_i-\mu\bigr).
\end{align*}
 Consequently, the marginal posteriors of $\gamma$ and $K$ are also normal. This structure enables understanding the analytic form of the posterior which is helpful for potential custom implementations tailored to our estimator. Finally, the marginal posterior of $\mu$ given the other 
parameters is normal as well,
so that a Gibbs 
sampler may be implemented to obtain a sample from the posterior and, consequently, from the 
predictive distribution of the targeted posterior missing labels, as we later specify.



\section*{Closed-form for one-dimensional posterior}

Up to an additive constant that does not depend on the parameters, the log-posterior for a single-source model is
$$\begin{aligned}
	\log p(\gamma,K,\mu,\sigma_Y^2,\sigma_X^2,\sigma_\mu^2,\tau_Y^2,\tau_\mu^2\mid D_1)
	\;&\propto\;
	-\frac{n}{2}\log \sigma_Y^{2}
	-\frac{1}{2\sigma_Y^{2}}\sum_{i=1}^{n}\!\left(Y_i-\gamma X_i-K\bigl(X_i-\mu\bigr)\right)^{2}\\
	&\quad
	-\frac{n}{2}\log \sigma_X^{2}
	-\frac{1}{2\sigma_X^{2}}\sum_{i=1}^{n}(X_i-\mu)^2\\
	&\quad
	-\frac{1}{2}\log\!\bigl(\tau_Y^{2}\sigma_Y^{2}\bigr)
	-\frac{\gamma^{2}}{2\,\tau_Y^{2}\sigma_Y^{2}}
	-\frac{1}{2}\log\!\bigl(\tau_Y^{2}\sigma_Y^{2}\bigr)
	-\frac{K^{2}}{2\,\tau_Y^{2}\sigma_Y^{2}}\\
	&\quad
	-\frac{1}{2}\log\!\bigl(\tau_\mu^{2}\sigma_\mu^{2}\bigr)
	-\frac{(\mu-\hat\mu)^{2}}{2\,\tau_\mu^{2}\sigma_\mu^{2}}\\
	&\quad
	-(a_Y+1)\log\sigma_Y^{2}-\frac{b_Y}{\sigma_Y^{2}}
	\;-\;(a_X+1)\log\sigma_X^{2}-\frac{b_X}{\sigma_X^{2}}\\
	&\quad
	-(a_\mu+1)\log\sigma_\mu^{2}-\frac{b_\mu}{\sigma_\mu^{2}}\\
	&\quad
	-\bigl(a_{\tau_Y}+1\bigr)\log\tau_Y^{2}-\frac{b_{\tau_Y}}{\tau_Y^{2}}
	\;-\;\bigl(a_{\tau_\mu}+1\bigr)\log\tau_\mu^{2}-\frac{b_{\tau_\mu}}{\tau_\mu^{2}}.
\end{aligned}
$$

\end{document}